\documentclass[journal=jacsat,manuscript=article]{achemso}
\usepackage[version=3]{mhchem} 
\usepackage{soul}

\usepackage{tablefootnote}
\usepackage{threeparttable} 

\author{Gonzalo Díaz Mirón}
\affiliation[ICTP]{Condensed Matter and Statistical Physics, The Abdus Salam International Centre for Theoretical Physics, 34151 Trieste, Italy}
\email{gdiaz_mi@ictp.it}
\author{Carlos R. Lien-Medrano}
\affiliation[Bremen]{Institute for Theoretical Physics and Bremen Center for Computational Materials Science, University of Bremen, 28359 Bremen, Germany}
\author{Debarshi Banerjee}
\affiliation[ICTP]{Condensed Matter and Statistical Physics, The Abdus Salam International Centre for Theoretical Physics, 34151 Trieste, Italy}
\alsoaffiliation[SISSA]{Scuola Internazionale Superiore di Studi Avanzati (SISSA), 34136 Trieste, Italy}
\author{Uriel N. Morzan}
\affiliation[DFUBA]{Instituto de Fisica de Buenos Aires, Facultad de Ciencias Exactas y Naturales, Universidad de Buenos Aires, C1428EGA Buenos Aires, Argentina}
\author{Michael A. Sentef}
\affiliation[Bremen]{Institute for Theoretical Physics and Bremen Center for Computational Materials Science, University of Bremen, 28359 Bremen, Germany}
\alsoaffiliation[MPSD]{Max Planck Institute for the Structure and Dynamics of Matter, Center for Free-Electron Laser Science (CFEL), 22761 Hamburg, Germany}
\author{Ralph Gebauer}
\affiliation[ICTP]{Condensed Matter and Statistical Physics, The Abdus Salam International Centre for Theoretical Physics, 34151 Trieste, Italy}
\author{Ali Hassanali}
\affiliation[ICTP]{Condensed Matter and Statistical Physics, The Abdus Salam International Centre for Theoretical Physics, 34151 Trieste, Italy}
\email{ahassana@ictp.it}

\title{Exploring the Mechanisms Behind Non Aromatic Fluorescence with the Density Functional Tight Binding Method}

\abbreviations{NAF,DFTB,NAMD}

\SectionNumbersOn

\begin{document}

\begin{tocentry}
\begin{figure}[H]
  \centering
  \includegraphics[width=1.0\linewidth]{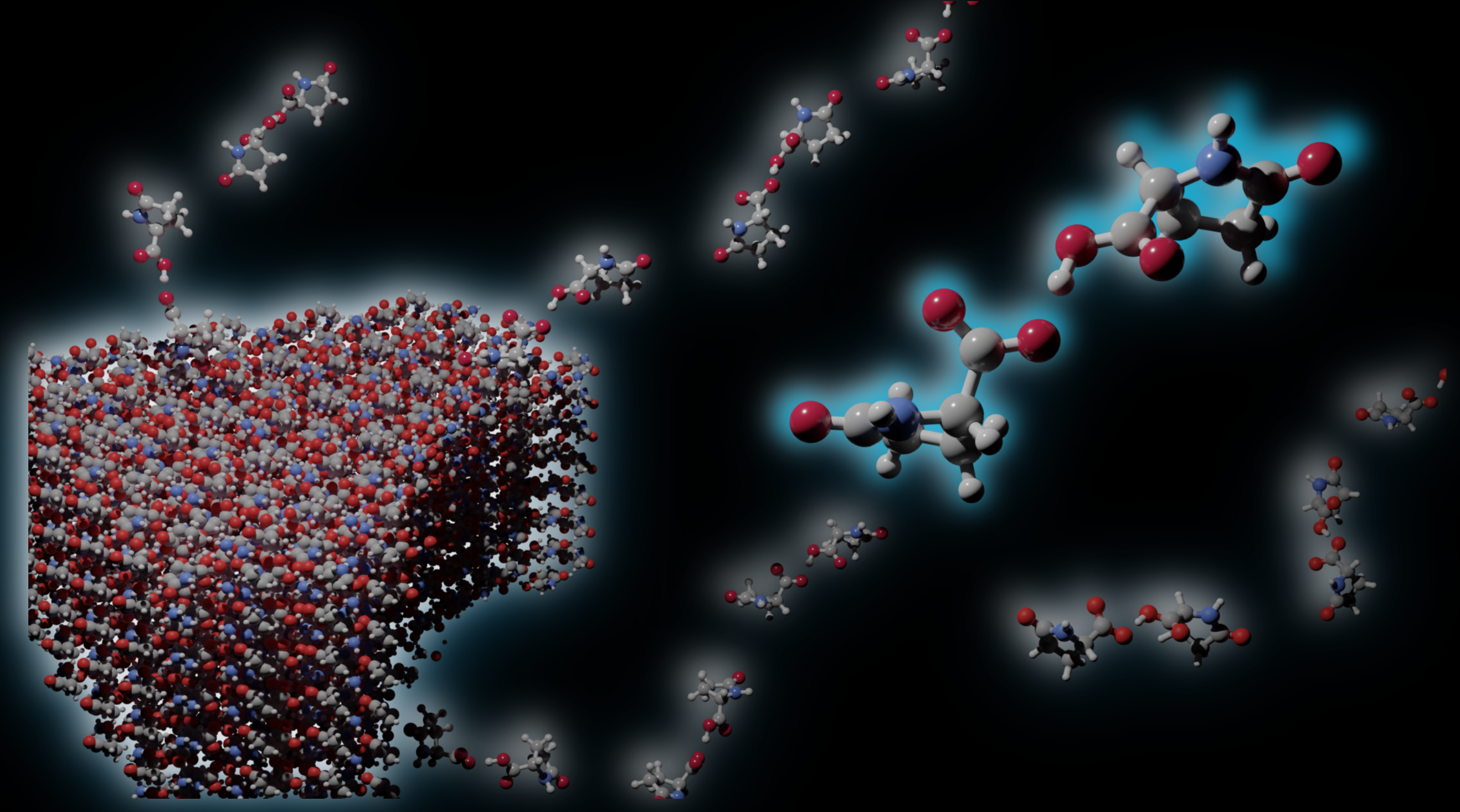}
  \label{fig:toc}
\end{figure}
\end{tocentry}

\newpage
\begin{abstract}
Recent experimental findings reveal non-conventional fluorescence emission in biological systems devoid of conjugated bonds or aromatic compounds, termed \textit{Non-Aromatic Fluorescence} (NAF). This phenomenon is exclusive to aggregated or solid states, remaining absent in monomeric solutions. Previous studies focused on small model systems in vacuum show that the carbonyl stretching mode along with strong interaction of short hydrogen bonds (SHBs) remain the primary vibrational mode explaining NAF in these systems. In order to simulate larger model systems taking into account the effects of the surrounding environment, in this work we propose using the density functional tight-binding (DFTB) method in combination with non-adiabatic molecular dynamics (NAMD) and the mixed quantum/molecular mechanics (QM/MM) approach. We investigate the mechanism behind NAF in the crystal structure of L-pyroglutamine-ammonium, comparing it with the related non-fluorescent amino acid L-glutamine. Our results extend our previous findings to more realistic systems, demonstrating the efficiency and robustness of the proposed DFTB method in the context of NAMD in biological systems. Furthemore, due to its inherent low computational cost, this method allows for a better sampling on the non-radiative events at the conical intersection which is crucial for a complete understanding of this phenomenon. Beyond contributing to the ongoing exploration of NAF, this work paves the way for future application of this method in more complex biological systems such as amyloid aggregates, biomaterials and non-aromatic proteins.
\end{abstract}

\newpage
\section{Introduction}

Over the last decade, a growing body of experimental evidence has revealed the emergence of non-conventional fluorescence signals in biomolecular systems. In contrast to conventional spectroscopic intuition attributing fluorescence in biological systems to aromatic or conjugated groups, recent experiments have shown that non-aromatic systems exhibit inherent absorption and fluorescence in the UV-visible range\cite{morzan2022non,ma2021bright,shukla2004novel,del2007charge}. These fluorescent signals appear to be enhanced in aggregated forms, analogous to the \textit{aggregation-induced emission} observed in organic materials\cite{tang2021nonconventional,zhou2016clustering}. Noteworthy examples include amyloid structures\cite{pinotsi2013label,pansieri2019ultraviolet,grisanti2020toward}, individual amino acids\cite{arnon2021off,homchaudhuri2001novel}, compounds derived from amino acids\cite{stephens2021short,arnon2021off}, poly-amidoamines\cite{tsai2011intrinsically} and small organic molecules\cite{guan2020strategy,fang2018unexpected}. 

While the detailed molecular mechanism underlying \textit{Non-Aromatic Fluorescence} (NAF) remains an open question, several possible explanations have been proposed\cite{morzan2022non,tang2021nonconventional}. These include electron delocalization along peptide bonds in ordered secondary structures\cite{shukla2004novel}, the constriction of the carbonyl stretching mode caused by intense local interactions such as strong hydrogen bonds\cite{miron2023carbonyl,stephens2021short}, proton transfer along short hydrogen bonds\cite{pinotsi2016} and inter-residue charge-transfer excitations followed by charge recombination\cite{del2007charge}. These molecular phenomena are believed to introduce low lying fluorescent electronic states in the visible regime arising from the supramolecular assemblies that do not rely on the presence of $\pi$-delocalized electrons. Since fluorescence emission involves the de-activation from electronic excited states, it requires a complex interplay involving the dynamics of both nuclear and electronic degrees of freedom which are notoriously difficult to model.

Over the last few years, our group has been leading theoretical and computational efforts to investigate the spectroscopic origins of this phenomenon\cite{grisanti2020toward,miron2023carbonyl,stephens2021short,grisanti2017computational,morzan2022non}. We have recently investigated NAF in model amyloid proteins\cite{grisanti2020toward,grisanti2017computational} and L-glutamine/L-glutamine-derived\cite{miron2023carbonyl,stephens2021short} amino-acids combining both experiments and theory. Employing trajectory surface hopping (TSH)\cite{tully1990molecular,barbatti2011nonadiabatic} simulations in combination with time-dependent density functional theory (TD-DFT) level\cite{casida2009time}, we have shown that carbonyl groups (CO) appear to play a critical role in modulating the observation of NAF in a wide variety of systems. Upon an initial photo-excitation to the first excited state $S_1$, which is mainly localized around the CO group, strong inter-molecular interactions involving hydrogen bonds within certain biological aggregates hinder the stretching of the CO bond. This leads to a restriction of the non-radiative relaxation pathways, thus boosting the fluorescence quantum yield. The CO stretching mode is strongly coupled to the proton transfer and deplanarization modes in the amide group, which are also restrained due to environment effects\cite{miron2023carbonyl}.

Our previous research has been based on dimer model systems carved out of crystal structures; in these studies we have accounted for the steric effect of the surrounding molecules by incorporating specific constraints in our simulations\cite{stephens2021short,grisanti2020toward,miron2023carbonyl}. The computational demands associated with simulating the photophysics of these systems make TSH calculations extremely challenging. Furthermore, the inherent variability in the electronic relaxation pathways can introduce problems with convergence. Consequently, the inclusion of a substantial number of trajectories becomes essential to derive meaningful and reliable results. We have recently employed hybrid Quantum-Classical Mechanical (QM/MM) approaches to enable a realistic description of environmental effects on the fluorescence.\cite{miron2023carbonyl} However, these approaches are computationally limited by the size of the relevant quantum-mechanical subsystem. These limitations have sparked our interest in exploring alternative methods to effectively tackle these challenges. In this regard, the density functional tight-binding (DFTB) method\cite{Cui2014,Elstner1998}, a semi-empirical approach based on density functional theory (DFT), has emerged as a promising avenue. DFTB achieves computational efficiency through a combination of the two-center approximation, pre-tabulated integrals, and the use of a minimal basis set\cite{Koskinen2009,elstner2014density,gaus2014density}. This accelerates DFTB calculations by a factor of 2-3 order of magnitude compared to traditional DFT\cite{elstner2014density}. Moreover, the pre-tabulation of integrals at DFT level ensures the precision of DFTB in predicting electronic properties for molecular, nano, and periodic systems, effectively emulating DFT\cite{Gaus2013dftb,Gruden2017dftb,Poidevin2023dftb,Fihey2015dftb,Maghrebi2023dftb,Dolgonos2010dftb}. This framework has also been extended to include the modeling of excited state dynamics leading to the time-dependent density functional tight-binding method (TD-DFTB). These extensions include both the frequency\cite{niehaus2009approximate,niehaus2001tight}(TD-DFTB) and the time domain\cite{bonafe2020real,uratani2021scalable} (real-time TD-DFTB). In addition, TD-DFTB has been coupled to the Ehrenfest approach\cite{bonafe2020real,uratani2021scalable} and TSH\cite{stojanovic2017nonadiabatic,mitric2009nonadiabatic,posenitskiy2019non} to perform non-adiabatic dynamics with semiclassical ion motion.

The goal of the present work is to comprehensively assess the potential of the TD-DFTB method in investigating the photophysics underlying NAF phenomenon employing the DFTB+ code\cite{hourahine2020dftb+}. The primary focus of our investigation is on amino-acid crystals which have recently been studied in our group using both \textit{ab-initio} electronic methods (TD-DFT) and experiments. It was found that L-glutamine, characterized by regular hydrogen bond interactions, is a non-fluorescent system. However, following a chemical transformation into L-pyroglutamine-ammonium (L-pyro-ammonium, pyroglutamic acid complexed with an ammonium ion), the system undergoes a transition to a short hydrogen bond configuration and displays a remarkable fluorescent signal. Panels a and b of Figure \ref{fig:modes_pbc} show the molecular structure of these systems. By comparing and contrasting the mechanisms obtained with TD-DFTB to our previous work, we demonstrate that the DFTB approach correctly captures the non-radiative decay pathways occurring in L-glutamine as well as the origins of fluorescence in L-pyro-ammonium. With this validation in hand, we use TD-DFTB to understand the role of environmental effects using a QM/MM approach allowing for an accurate characterization of the experimentally observed Stokes shift between the excitation and emission spectra in this system. We also highlight aspects of non-radiative decay mechanisms which are different from our previous studies paving the way for future directions and developments needed to use Tight-Binding approaches to study non-aromatic fluorescence.

\section{Theoretical Approaches}

The objective of this section is to provide a background on the various theories and techniques employed in our methodologies to study non-aromatic fluorescence. In this section, we will summarize some of the most important aspects of DFTB, as well as its Time-Dependent extension to the frequency domain (TD-DFTB), with a special focus on the implementations in the DFTB+ code\cite{hourahine2020dftb+}. Additionally, we will describe the key aspects of non-adiabatic dynamics in the context of Trajectory Surface Hopping (TSH), with a specific emphasis on the different analytical expressions used to calculate the non-radiative probability decay. 

\subsection{Density Functional Tight-Binding (DFTB and TD-DFTB)}

DFTB equations are derived from DFT by expansion of the total energy in a Taylor series of the electron density fluctuations $\delta\rho$ around a reference density $\rho^0$\cite{hourahine2020dftb+}. 
The chosen reference density is usually a summation of overlapping, spherical and noninteracting atomic charge densities. 
Up to the third order, the total energy can be approximated as:\cite{Gaus2011}

\begin{equation}
    E^{\text{DFT}}[\rho^0+\delta\rho] \approx E^0[\rho^0] + E^{1\text{st}}[\rho^0,\delta\rho] + E^{2\text{nd}}[\rho^0,(\delta\rho)^2] + E^{3\text{rd}}[\rho^0,(\delta\rho)^3]. 
    \label{eq:tbdft}
\end{equation}
The initial term $E^0$ can be represented by summing pairwise repulsive potential energies $E^{rep}_{AB}$. The truncation up to first order in the energy can be expressed as

\begin{equation}
    E^{\text{DFTB1}} = E^0 + E^{1\text{st}} =  \sum_{A>B}E_{AB}^{\text{rep}} + \sum^{\text{occ.}}_i n_i \langle{\psi_i}|\mathcal{H}[\rho^0]|\psi_i\rangle,
    \label{eq:dfbt1}
\end{equation}
where $|\psi_i\rangle$ is the $i^{\text{th}}$ molecular orbital in a linear combination of atomic orbitals (LCAO), $n_i$ is the occupation and $\mathcal{H}[\rho^0]$ is the Hamiltonian operator in a two-center approximation. 

In the second-order DFTB (DFTB2) method, the $E^{2\text{nd}}$ term is incorporated through a monopole approximation\cite{Elstner1998}, where the electron density fluctuations are approximated as a sum of atomic Slater-type spherical charge densities:

\begin{equation}
    E^{\text{DFTB2}} = E^{\text{DFTB1}} + \frac{1}{2} \sum_{AB} \gamma_{AB} \Delta q_A \Delta q_B.
    \label{eq:dfbt2}
\end{equation}
Here, $\Delta q_A$ is the Mulliken charge on atom $A$ and $\gamma_{AB}$ represents the electron interaction of two Slater-type spherical charge densities on atoms $A$ and $B$. The inclusion of the second-order term requires a self-consistent solution since the Mulliken charges depend on the molecular orbitals. This provides a more accurate description of charge-transfer and charge-charge interactions. 

DFTB3 allows the chemical hardness of an atom to change with its charge states by including third order terms:
\begin{align}
    E^{\text{DFTB3}} = E^{\text{DFTB2}} + \frac{1}{3} \sum_{AB}\Gamma_{AB}\Delta q_A^2 \Delta q_B,
    \label{eq:dfbt3}
\end{align}
where $\Gamma_{AB}$ is the derivative of $\gamma_{AB}$ with respect to the $\Delta q_A$.

Corrections for both van-der-Waals\cite{brandenburg2014accurate,miriyala2017description} and hydrogen-bonding interactions\cite{rezac2017empirical,rezac2009semiempirical} were also implemented in the method owing to the fact that these types of weak interactions are not accurately captured by DFT based methods including DFTB.

Within the DFTB+\cite{hourahine2020dftb+} code there are several available approaches to determine excited states. We employed the Casida formalism\cite{casida2009time} within the framework of DFTB, following the same ansatz as in the original version of TD-DFT\cite{niehaus2001tight,niehaus2009approximate}. 
The electronic excitation energy $\omega_{I}$ in TD-DFTB can be obtained by solving the following eigenvalue problem:

\begin{equation}
   \displaystyle \Omega\mathbf{F}_I = \omega^2_I\mathbf{F}_I.
    \label{eq:casida}
\end{equation}
Here, $\Omega$ is the response matrix and its elements are
\begin{equation}
  \displaystyle  \Omega_{ij\sigma,kl\tau}= \delta_{ik}\delta_{jl}\delta_{\sigma\tau}\omega^2_{ij}
    +2\sqrt{\omega_{ij}}K_{ij\sigma,kl\tau}\sqrt{\omega_{kl}},
  \label{eq:omega}
\end{equation}
where $\omega_{ij}=\epsilon_j-\epsilon_i$, $i,k$ and $j,l$ are occupied and unoccupied KS orbitals respectively, $\sigma$ and $\tau$ are spin indices, whereas $K$ is the so-called coupling matrix. Within the monopole approximation, the coupling matrix elements $K_{ij\sigma,kl\tau}$ adopt simple expressions, reducing the computational cost.\cite{niehaus2009approximate} 

Moreover, the incorporation of Long-range Corrections (LC) into the two-electron integrals have also been implemented in the DFTB+ code\cite{niehaus2012range} with the aim to provide a reduction in the self-interaction error in DFT (and also present in DFTB) yielding a correct behaviour of the asymptotic potential. This implementation along with a new reparametrization of the Slater-Koster parameters enables a more accurate description of the excited states, mainly for those which involves charge transfer excitations\cite{lutsker2015implementation, ob2SKF2018}. It is worth noting that in previous literature, the term TD-LC-DFTB denotes the incorporation of LC into TD equations. However, in our current work, we will use the term TD-DFTB since all calculations within this theory were conducted with LC.

\subsection{Trajectory Surface Hopping (TSH)}

Trajectory Surface Hopping (TSH) is one of the most widely employed techniques to model non-adiabatic dynamics\cite{barbatti2011nonadiabatic}. The non-radiative relaxation is a non-equilibrium process and  therefore the dynamics strongly depends on the initial conditions. For this reason, the simulation of a large number of thermally distributed independent trajectories, with varying initial conditions becomes essential\cite{subotnik2016understanding,subotnik2011decoherence}. In each TSH trajectory, the electronic states are propagated using quantum mechanics, while the nuclear motion is handled classically employing the forces coming from a single potential energy surface from the quantum method. For more information on the theoretical and practical details, the reader is referred to previous studies\cite{barbatti2011nonadiabatic,crespo2018recent}. Bellow, we highlight specific details relevant to our present work.

The TSH workflow involves propagating the system on distinct potential energy surfaces. At each timestep, the non-radiative probability is computed and a stochastic algorithm is employed to decide which potential energy surface the system will proceed along. The most common expression for the transition probability is the Fewest Switches Surface Hopping (FSSH) proposed by Tully\cite{tully1990molecular}:

\begin{equation}
    P_{ij}(t) = -2 \int_t^{t+\delta t} dt' \dfrac{C_i(t')C_j(t')^*g_{ij}(t')}{C_i(t')C_i(t')^*},
    \label{eq:tully}
\end{equation}
where $C_i$ are the electronic coefficients of the potential energy surfaces $i$ and the $g_{ij}$ terms are the non adiabatic coupling (NAC) elements, one of the key components in non-adiabatic dynamics. Several alternatives approaches have been developed in order to avoid the explicit determination of the non adiabatic coupling elements\cite{nikitin1968theory,hayashi2009photochemical,zhu1993two}, since sometimes there are no analytical expressions available or because they are computationally expensive to determine for large systems. One of the most popular in this context is the Landau-Zener (LZSH) approximation\cite{zener1932non}:

\begin{equation}
    P_{ij}(t) = \exp{\left( - \dfrac{\pi}{2\hbar} \sqrt{\dfrac{\Delta E_{ij}^3(t)}{\dfrac{d^2}{dt^2}\Delta E_{ij}(t)}} \right)},
    \label{eq:landau}
\end{equation}
where $\Delta E_{ij}$ is the adiabatic energy gap\cite{xie2019assessing,crespo2018recent}. Within the context of our TD-DFTB simulations, we employ the LZSH approximation instead of Tully's method due to its easy implementation into our computational framework.

\section{Computational Details}

In this section, we summarize the computational protocols applied to L-glutamine (L-gln) and L-pyroglutamine ammonium (L-pyro-amm) to simulate the ground-state molecular dynamics in periodic boundary conditions, the determination of the absorption spectra, the excited states surface hopping dynamics and finally the QM/MM simulations with DFTB and TD-DFTB. In the present study, we employed the DFTB+ package\cite{hourahine2020dftb+} for the electronic structure calculations. The Genesis\cite{kobayashi2017genesis,jung2015genesis} software was used to perform classical and QM/MM simulations.

\subsection{Ground State DFTB Molecular Dynamics} \label{sec:gs_pbc}

\textit{Ab-initio} molecular dynamics simulations of both L-gln and L-pyro-amm were conducted using DFTB in the ground-state with periodic boundary conditions (PBC) using the gamma point. The initial coordinates for both systems were obtained from previously resolved crystallographic structures (CSD 1169316\cite{crystalglutamine} and 1981551\cite{stephens2021short}). We employed third order corrections in the Hamiltonian with the corresponding Slater Koster parameters 3ob\cite{param3ob} as well as D3 Grimme corrections\cite{brandenburg2014accurate}. Molecular dynamics simulations were conducted for a total of 600 ps for each system using a timestep of 0.5 fs and within the NVT ensemble using the Berendsen thermostat with a coupling of 0.1 ps$^{-1}$. For all the analysis, the first 20 ps were discarded as equilibration. As will be shown later, we compare some of the ground-state structural properties obtained from DFTB to those obtained with DFT. Details of these simulations can be found in the Computational Methods Section of a previous work in our group\cite{stephens2021short}.

\subsection{Absorption Spectra} 
\label{sec:abs_pbc}

After conducting the DFTB simulation in the ground state for L-gln and L-pyro-amm, we proceeded to extract frames at intervals of 500 fs. A total of 1000 frames were used to compute the absorption spectra for each system with the aim to allow for adequate structural fluctuations in the crystal and their subsequent coupling to electronic transitions.

These calculations were carried out within the TD-DFTB method without PBC, which corresponds to the simulation of a cluster of the unit cell for each system surrounded by vacuum. To account for the correct description of the excited states, we also incorporated Long-range Corrections (LC) with the appropriate Slater-Koster files\cite{lutsker2015implementation, ob2SKF2018}. A total of 100 states were calculated for each snapshot employing the Casida algorithm. The final absorption spectrum for each system was generated by averaging the frames, and each individual transition was broadened using a Gaussian function with a Full Width at Half Maximum (FWHM) of 5 nm.

\subsection{Non Adiabatic Molecular Dynamics in Gas Phase}
\label{sec:tsh_met}

TSH simulations were performed on dimer model systems carved out from the crystal structures of L-gln and L-pyro-amm. The dimer models were carefully selected to preserve the Short Hydrogen Bonds (SHBs) in L-pyro-amm and the normal Hydrogen Bonds (HBs) in L-gln. These simulations were conducted with the dimer in the gas phase. To account for the crystallographic environment, we applied atom constraints on specific selected atoms (see Figure S1 in the Supplementary Information).

Our protocol to generate the initial conditions consists of performing 6 ns long classical molecular dynamics simulations using the Generalized Amber Force Field (GAFF)\cite{wang2004development} for both model systems with a timestep of 1 fs and a thermostat constant of 1 ps$^{-1}$. From these long simulations, a total of 600 configurations were selected with a spacing of 10 ps. For each of these 600 configurations, ground-state simulations were continued using DFTB as described earlier in section \ref{sec:gs_pbc}, for 2 ps each at 300K. TSH simulations were started with the system in the electronic state $S_1$, corresponding to the initial photo-absorption, employing the positions and velocities of the previous step. The dynamics were then performed in the NVE ensemble for 250 fs using a timestep of 0.5 fs with TD-DFTB as described in section \ref{sec:abs_pbc}. At each timestep, the probability of non-radiative decay was calculated using the LZSH algorithm (Equation \ref{eq:landau}).

As noted earlier in the Introduction, one of the goals of this work is to compare and contrast the underlying non-radiative decay mechanisms associated with non-aromatic fluorescence obtained with TD-DFTB to the one reported in our earlier work\cite{miron2023carbonyl}, employing TD-DFT with the functional PBE0 and 6-31G as the basis set.

It is therefore important to recall two important differences: (\textit{i}) the use of TD-DFTB expands the statistics of TSH trajectories by a factor of 3 (600 vs 200) in the case of L-gln and a factor of 6 for L-pyro-amm (600 vs 100), (\textit{ii}) as already indicated earlier, the TD-DFTB hopping events between different states are performed using LZSH (Equation \ref{eq:landau}) while in our previous TD-DFT simulations, were conducted with FSSH (Equation \ref{eq:tully}). Despite this methodological difference, the comparison between the two approaches remains valid, as both theories have been extensively tested and compared in several works\cite{xie2020performance,xie2017accuracy,suchan2020pragmatic}, demonstrating consistent results.

\subsection{QM/MM Simulations}
The accurate simulation of the photophysical relaxation of the excited states in these crystals require the consideration of PBC. However, DFTB+ code does not calculate forces in the excited states within these conditions. One potential solution to circumvent this challenge is to conduct simulations on clusters in the gas phase, as discussed in the preceding section. Another more effective alternative is to employ QM/MM simulations to elucidate the influence of the environment\cite{crespo2003dft,senn2009qm}. This method involves partitioning the system into two distinct regions, namely a QM and MM, and describing the coupling between these regions using the electrostatic embedding approach\cite{morzan2018spectroscopy,senn2009qm,boulanger2018qm}. To investigate the radiative and non-radiative process in both systems, we selected two different QM sub-systems for L-pyro-amm, the dimer and unit cell, while for L-gln we studied the unit cell. 

To construct the system, we began by replicating the unit cell of the crystal structure in all three directions resulting in a supercell comprising 375 unit cells for both L-gln and L-pyro-amm. The subsequent step involved a classical equilibration at room temperature for 10 ns with a timestep of 1 fs. Classical parameters for bonded and Lennard-Jones were adopted from GAFF\cite{wang2004development} and for treating the electrostatics we used Mulliken charges which were extracted from optimized crystal structures using DFTB in PBC (see section \ref{sec:gs_pbc}). A total of 10 snapshots were extracted from the simulations.
 
For each snapshot, we proceed to select the QM region situated at the center of our supercell, then a QM/MM ground state dynamics simulation of 25 ps each were conducted with the same simulation protocol. From the last 20 ps of these simulations, we extracted 20 frames for each trajectory to calculate the absorption spectra using TD-DFTB in the crystal environment (see section \ref{sec:abs_pbc}). A total of 200 snapshots were utilized for the absorption spectra.

In addition, we performed TSH simulations for each of the QM sub-system with the aim to study the emission as well as the photo-physical relaxation of the $S_1$ state of these systems.
TSH simulations started from the $S_1$ electronic state employing the initial conditions generated in the previous step for a total of 1 ps, with a timestep of 0.5 fs within the NVE ensemble. A total of 10 TSH dynamics were conducted for each system using the same methodology described section \ref{sec:tsh_met}. For the fluorescence spectra, we employed trajectories that did not show decay to the ground state and excluded the initial 200 fs from the analysis, which corresponds to the time for the system to reach the minimum on $S_1$. The emission spectra were computed by utilizing the energy gap between $S_1$ and $S_0$, along with the oscillator strength, and then averaging.

\section{Results and Discussion}

\subsection{Ground State Properties in Molecular Crystals}

\begin{figure}[H]
  \centering
  \includegraphics[width=0.9\linewidth]{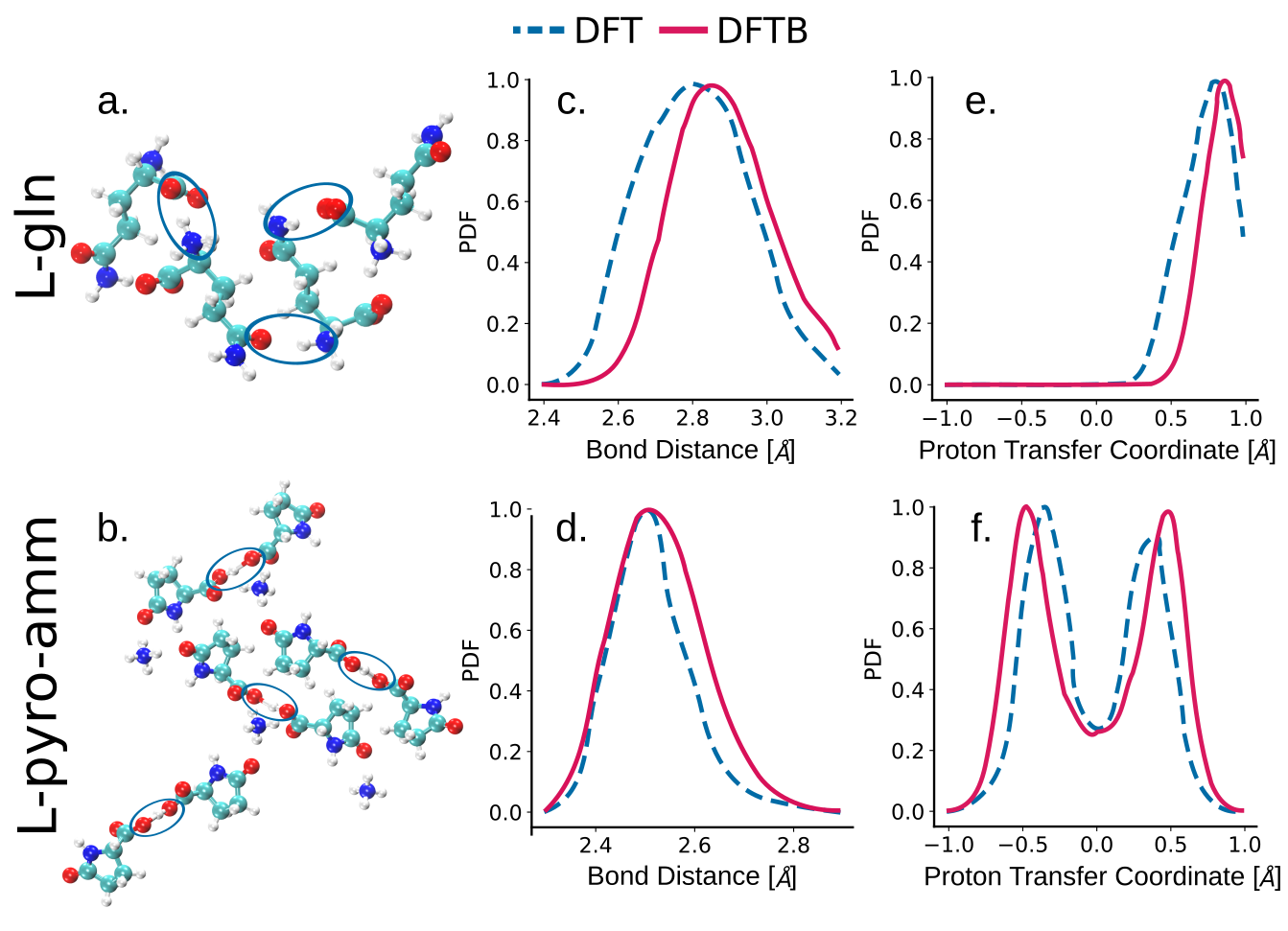}
  \caption{Unit cell crystallographic geometries for L-gln (panel a) and L-pyro-amm (panel b), where the blue circles highlight the HBs and SHBs, respectively. Histograms of BD (panels c and d) and PTC (panels e and f) obtained from an \textit{ab-initio} ground state dynamics with PBC using DFT (blue dashed line) and DFTB (red solid lines)}
  \label{fig:modes_pbc}
\end{figure}

Before analyzing the optical properties of these non-aromatic systems with TD-DFTB, it is necessary first to verify the quality of the hydrogen bonds (HBs) description, in particular the difference in the distance and its strength for the L-Pyroglutamine-Ammonium (L-pyro-amm) and L-Glutamaine (L-gln) in the crystal structures, since this property plays an important role in the fluorescence\cite{stephens2021short,grisanti2020toward}. Figure \ref{fig:modes_pbc} shows the probability distribution function (PDF) obtained from \textit{ab-initio} ground state molecular dynamics with PBC (see section \ref{sec:gs_pbc}) for the typical descriptors employed to describe the HBs including the proton transfer coordinate (PTC) and bond distance (BD) between the two heavy atoms among which proton transfer potentially occurs. BD is defined as the distance between O and N atoms in the case of L-gln (panel a and c) and the distance between the two O atoms in the case of L-pyro-amm (panel b and d). The PTC is defined as the distance difference between the O-H and N-H in L-gln (panel e) and between the two O-H involved in L-pyro-amm (panel f).

Overall, we observe that DFTB works very well at reproducing the properties of the HBs in both L-gln and L-pyro-amm. Specifically, the bond lengths associated with the SHBs are shorter than those of normal HBs comparing the two systems ($2.4-2.6$ \r{A} vs $ 2.7-3.0$ \AA). Another important feature of the SHBs in L-pyro-amm was the proton transfer, which has been observed in our previous DFT simulations\cite{stephens2021short}. This is reflected in a double-well potential along the PTC. Panels e and f show that DFTB accurately reproduces this feature for the SHBs and the single asymmetric well of the normal HBs. Given the nature of the DFTB method along with its various approximations, we find the agreement very encouraging. 

\subsection{Absorption Spectra in Molecular Crystals} \label{sec:resultsPBC_abs}

\begin{figure}[H]
  \centering
  \includegraphics[width=1.0\linewidth]{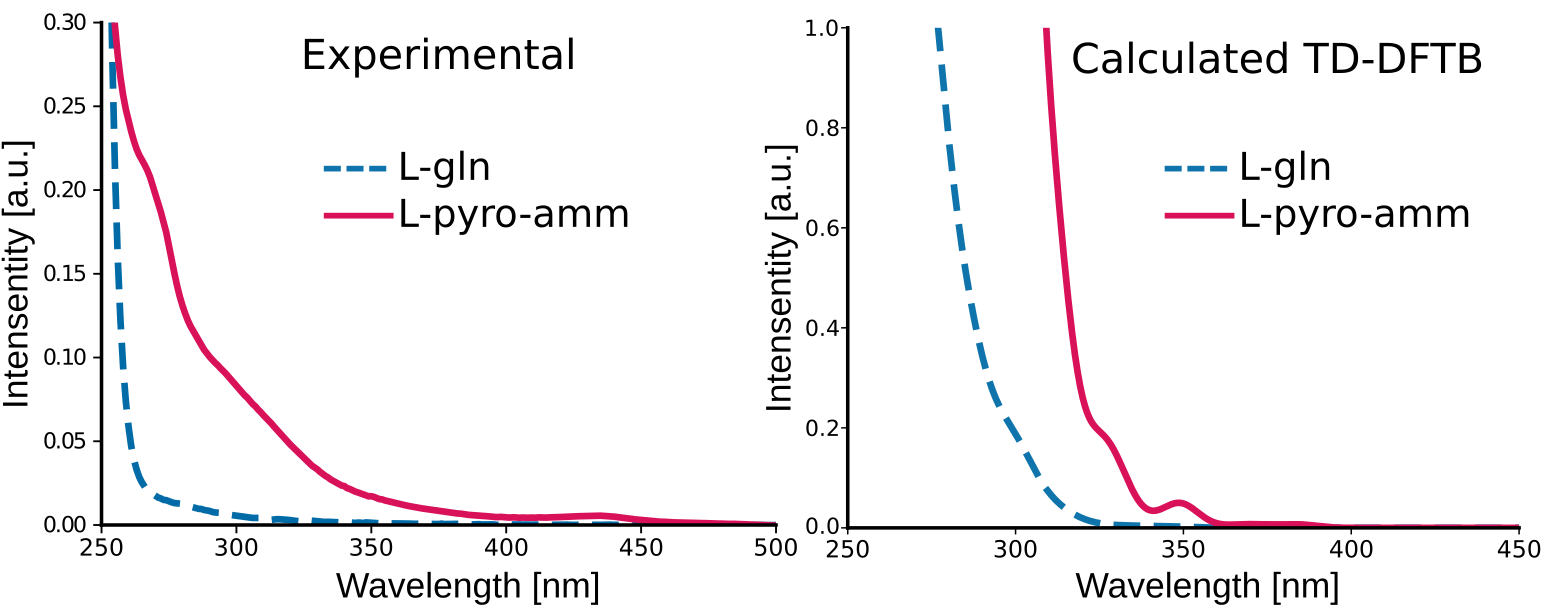}
  \caption{Absorption UV-Vis Spectra for L-Glutamine and L-Pyro-Ammonium. Left panel shows the experimental data, extracted from a previous study in our group\cite{stephens2021short}. Right panel shows the calculated average spectra using TD-DFTB. Red solid lines correspond to L-pyro-amm system while blue dashed lines for L-gln.}
  \label{fig:absPBC}
\end{figure}

With the ground-state configurations obtained from the \textit{ab-initio} MD, we next turn to extracting the absorption spectra. Figure \ref{fig:absPBC} shows the absorption spectra for L-gln and L-pyro-amm obtained experimentally\cite{stephens2021short} (left panel) as well as those calculated with TD-DFTB (right panel). As one can observe, an important difference between the two systems is the extent of the red-tail/edge excitation that is significantly more pronounced in the case of L-pyro-amm. In the case of L-gln, the absorption peaks are below the 250 nm and outside the window of experimentally detectable wavelengths. On other hand for L-pyro-amm, there is a very pronounced red-edge excitation tail ranging from 250-400 nm. 

The right panel of Figure \ref{fig:absPBC} depicts the average calculated absorption spectra obtained for both systems using 1000 averaged snapshots each, at the level of TD-DFTB (full-range spectra are shown in Figure S2 in the Supplementary Information). Interestingly, we observe that TD-DFTB reproduces the correct experimental trends comparing L-gln and L-pyro-amm. Analysing the transitions involved in the UV-Vis energy range (330-370 nm) we found that the main molecular orbitals that explain these transitions involve significant contributions coming from the carbonyl bonds and the SHB (see Figure S3 in the Supplementary Information). It is also worth noting that this prediction is also fully consistent with the results that were previously obtained at TD-DFT level in our previous work\cite{stephens2021short}.

\subsection{Excited State TSH Simulations in Gas phase}

\begin{figure}[H]
  \centering
  \includegraphics[width=1.0\linewidth]{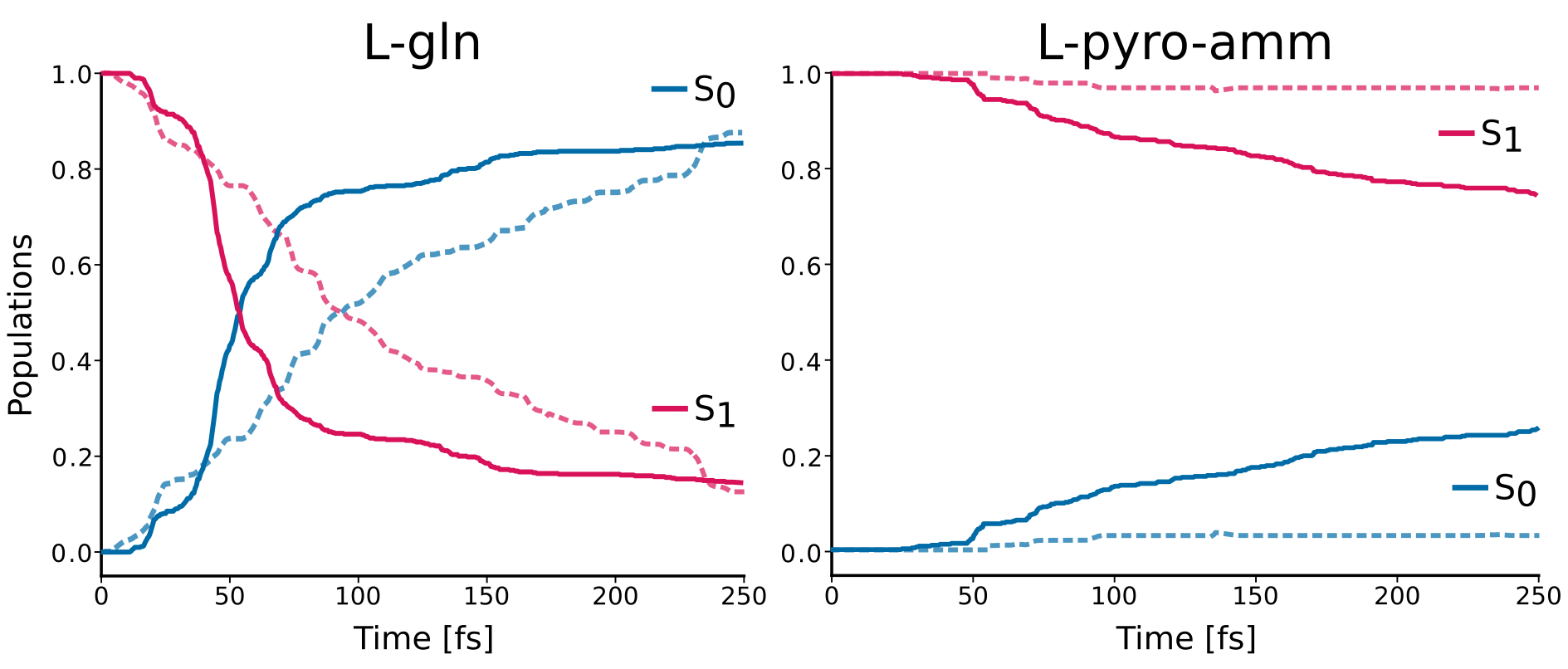}
  \caption{Average populations of the electronic states obtained from TSH trajectories for L-gln (left panel) and L-pyro-amm (right panel) dimer systems in vacuum. Results using TD-DFT are depicted as dashed lines while those obtained with TD-DFTB in solid lines. The evolution of $S_1$ and $S_0$ are denoted in red and blue lines, respectively.}
  \label{fig:populations}
\end{figure}

The two molecular systems (L-gln and L-pyro-amm) display substantially different photophysical properties. Specifically in L-pyro-amm, non-radiative decay pathways are effectively shut off\cite{miron2023carbonyl} thereby enhancing the excited state lifetime and subsequently increasing the fluorescence yield. From the TSH simulations on both systems, we extracted the average populations of the $S_1$ and $S_0$ electronic states over time which are illustrated in Figure \ref{fig:populations} for the dimer systems in vacuum. The solid lines come from the TD-DFTB simulations in this work, while the dotted lines are obtained from our previous TD-DFT calculations\cite{miron2023carbonyl}. 
From the evolution of the electronic states, we can observe that TD-DFTB captures the essential photophysics of both systems. In L-gln dimers, an ultra-fast non-radiative decay is evident, with a characteristic lifetime of $\sim$90 fs for TD-DFT and $\sim$50 fs with TD-DFTB. Conversely, L-pyro-amm exhibits a much longer $S_1$ lifetime with both theory levels. These findings are again, fully consistent with the experimental findings\cite{stephens2021short} where fluorescence is only observed in the L-pyro-amm molecular crystal. 

In our previous studies employing TD-DFT, we observed that the transition from $S_1\to S_0$ occurs via a conical intersection (CoIn) where the essential vibrational modes coupled to the process involves the extension of the carbonyl bond, proton transfer along the SHB, and finally, the deplanarization of the amide group. In this context, it is recognized that single-determinant methods such as TD-DFT (or TD-DFTB) present an inaccurate description of the $S_1/S_0$ CoIn. However, in our prior work (see Figure 5 in the SI in the reference\cite{miron2023carbonyl}), we validated the CoIn identified with TD-DFT using the CASPT2 method. We found very good agreement between both theory levels.

\begin{figure}[H]
  \centering
  \includegraphics[width=1.0\linewidth]{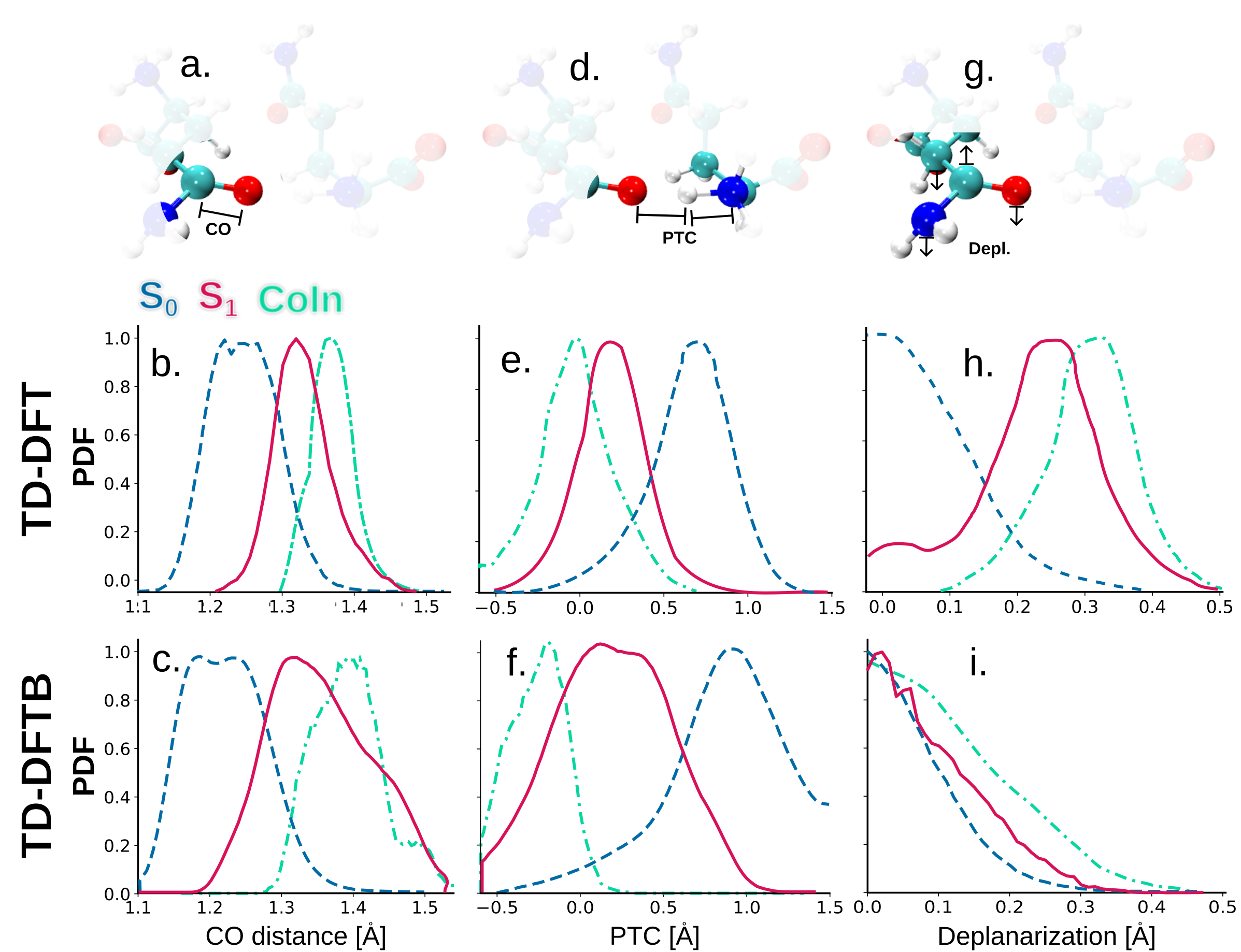}
  \caption{Structural features characterizing ground (blue dashed line) and excited states (red solid line) as well as CoIn (green dashed-dot line) crossing for L-gln system. Molecular structures showing the parameters to describe the carbonyl stretching (panel a), the proton transfer (panel d) and the deplanarization (panel g). Probability Distribution Function (PDF) for the CO distance (panels b and c), PTC (panels e and f) and Deplanarization modes (panels h and i) for L-gln dimers in vacuum obtained from TSH trajectories using TD-DFT (middle panels) and TD-DFTB (lower panels). All PDFs were normalized independently.}
  \label{fig:namd_gln}
\end{figure}

Examining each of the modes mentioned before in the ground and excited state as well as in the CoIn, a very specific structural signature becomes apparent. Figure \ref{fig:namd_gln} depicts the histograms for each of these modes obtained from TSH simulations employing both TD-DFT (top) and TD-DFTB (bottom) theory levels for L-gln dimer. The histograms were computed using data from all TSH trajectories. At each time step, we identified the current potential energy surface ($S_1$ or $S_0$), or if the system transitioned to the other surface (CoIn).

The comparison between TD-DFTB and TD-DFT reveals excellent agreement both in terms of the overall trends observed but also in the specific details of how the modes change during non-radiative decay. The variation observed in the maximum peak intensities of the PDFs for both $S_1$ and $S_0$ across all depicted modes in Figure \ref{fig:namd_gln} arises from the significant displacement between the minima on both potential energy surfaces. This displacement is a consequence of the substantial distortions experienced by the system upon excitation, a feature effectively captured by the DFTB approach. In particular, we observe that an additional stretching for CO respect to $S_1$ distribution is necessary for the system to reach the CoIn (green dashed-dot lines in panel b and c). Similar behavior is also observed for the PTC (panels e and f). The primary difference arises in relation to the deplanarization mode. In TD-DFT (panel h), we observe a higher degree of deplanarization at the CoIn compared to the distribution of this mode in the $S_0$ state. In contrast, although the distributions obtained with TD-DFTB reflect the same trends (panel i), the deplanarization mode appears to be much more constrained compared to our previous findings. Despite the differences, it should be underscored that, based in our previous work\cite{miron2023carbonyl}, the deplanarization mode is less critical for the photophysics of both L-pyro-amm and L-gln. 

We next turn to comparing the behavior of the non-radiative decay modes that are important for L-pyro-amm. Since this system is fluorescent, extracting the decay mechanisms that emerge from the excited state simulations is particularly challenging due to the fact that there are fewer statistics of reactive events. In our previous study\cite{miron2023carbonyl} for example, out of a total of 100 trajectories, 4 displayed non-radiative decay from the excited to the ground-state.

\begin{figure}[H]
  \centering
  \includegraphics[width=0.7\linewidth]{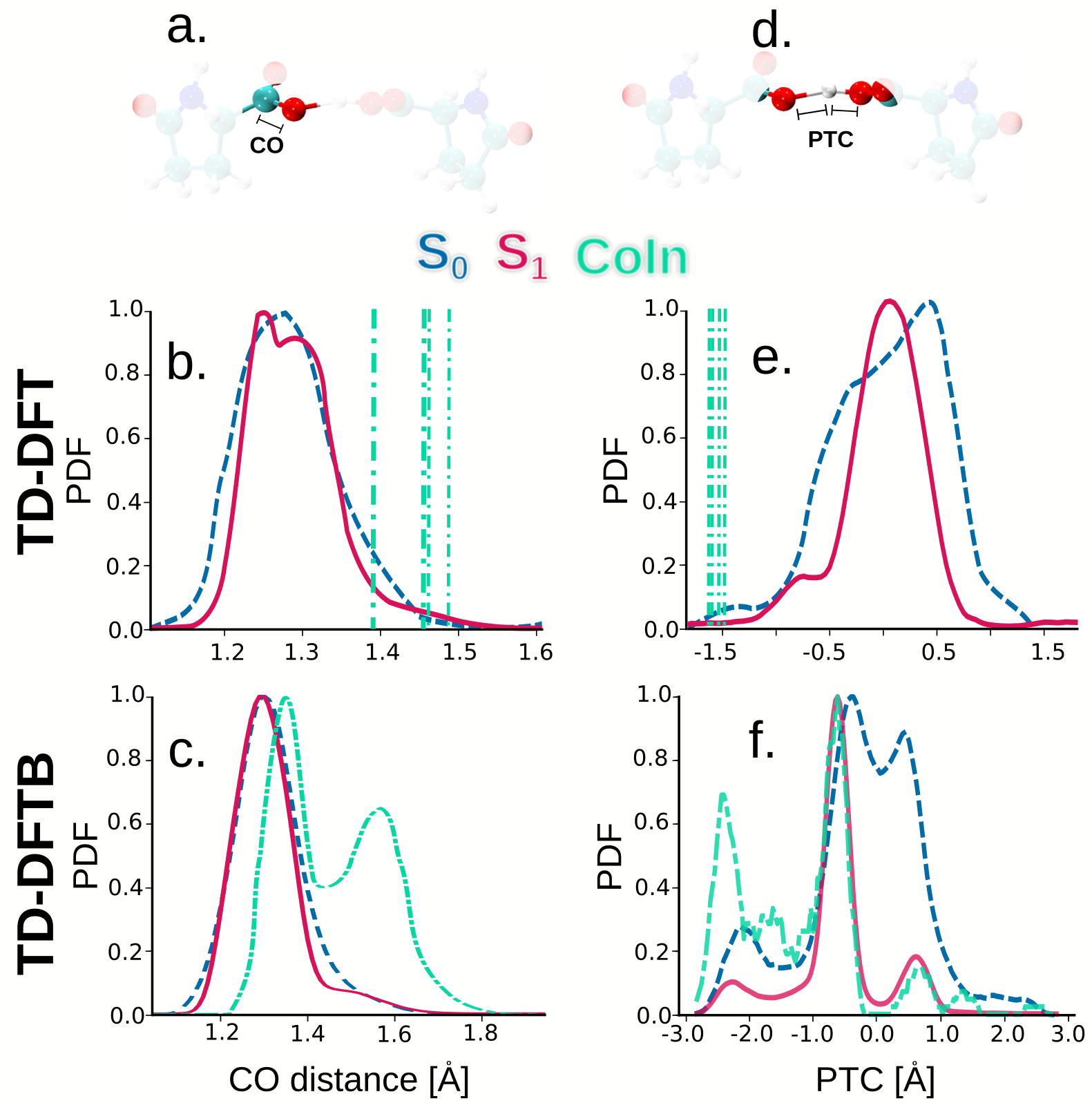}
  \caption{Structural features characterizing ground (blue dashed line) and excited states (red solid line) as well as CoIn (green dashed-dot line) crossing for L-pyro-amm system. Molecular structures showing the parameters to describe the carbonyl stretching and the proton transfer are in panel a and d, respectively. Probability Distribution Function (PDF) for the CO distance (panels b and c), PTC (panels e and f) for L-pyro-amm dimers in vacuum obtained from TSH trajectories using TD-DFT (middle panels) and TD-DFTB (lower panels). All PDFs were normalized independently.}
  \label{fig:namd_pyro}
\end{figure}

In Figure \ref{fig:namd_pyro} we show the most relevant features, namely the carbonyl stretching and the proton transfer modes obtained from the TSH simulations at TD-DFT and TD-DFTB levels. Specifically, we are able to reproduce the significant overlap in the distribution of the two modes between the $S_1$ and $S_0$ electronic states (red solid and blue dashed lines). These findings indicate that the energy minimum in the potential energy surfaces for each of these modes do not exhibit significant displacement, suggesting that these modes are restricted in L-pyro-amm. This feature is also fully consistent with typical aromatic fluorescent systems\cite{strickler1962relationship}, where the restriction arises from the high electron delocalization in the planar arrangement of these compounds. In the case of non-aromatic systems, like the ones we are tackling, the strong interactions due to SHBs induce this restriction on different modes.

It is important to highlight when comparing the top and bottom panels of Figure \ref{fig:namd_pyro}, that due to the reduced computational complexity, we are able to run significantly more TSH trajectories compared to our previous TDDFT simulations (600 vs 100). DFTB thus permits the observation of many more non-radiative decays yielding statistics which allow the construction of histograms of various modes near the CoIn, for instance the green distribution in the case of TD-DFTB represents 122 decays (panels c and f) and the green vertical lines for TD-DFT represents only 4 decays (panels b and e). The examination of the distribution of these modes at the CoIn reveals that the mechanism observed in L-pyro-amm is similar to that observed in the L-gln system. Specifically, we find with the DFTB approach that the CO stretching mode undergoes two distinct signatures at the CoIn. One exhibits a more subtle stretching in comparison to the distribution of $S_1$, while the other shows a more significant stretching of approximately $1.6$ \AA (green dashed-dot line in panel c in Figure \ref{fig:namd_pyro}). Our previous TD-DFT simulations sample a few events between these two peaks.

The effect of having more statistics with DFTB is observed more clearly in the PTC distributions (panels e and f) which display more structure compared to those obtained with TD-DFT. For example, we can capture the double-well character of the PTC (blue line in panel f) with TD-DFTB, while with TD-DFT we observe one main peak along with a shoulder (blue dashed line panel e). A distinction emerges when comparing the behavior of proton transfer between the excited and ground states in TD-DFTB. Notably, in the excited state, there is a significant reduction in proton transfer events, evidenced by the presence two peaks located at $\pm 0.5$ \r{A} with different intensity in the PDF (depicted by the red line in panel f). A similar trend is observed in TD-DFT, albeit with less intensity. In this case, a narrow distribution is noted in $S_1$ compared to the ground state, with the hydrogen positioned equidistant from the two oxygen atoms involved in the SHB (illustrated by the red line in panel e). Additionally, we observe that the SHB experiences a deformation at the CoIn (green dashed-dot line in panel f in Figure \ref{fig:namd_pyro}), transforming into a transient normal HB. This is evident from the peaks at higher values of approximately $\pm 2.5$ \AA, which is also present in TD-DFT (green dashed-dot lines in panel e). Although we cannot rule out the possibility that some of these differences arise from the underlying potential energy surface along various vibrational modes, the limited statistics with TD-DFT are more likely to be the source of the differences we observe. However, we can see that the mechanism predicted by TD-DFTB is in excellent agreement with our previous studies\cite{stephens2021short,miron2023carbonyl}.  

\subsection{Optical Properties in Realistic Biological Environments}

The preceding results provide evidence that DFTB is a very viable approach for studying the non-radiative decay mechanisms in organic molecular crystals implicated in non-aromatic fluorescence. As previously mentioned, to simulate a more realistic system beyond vacuum dimers, we adopted the QM/MM framework to account for environmental effects. Consequently, in the following section we utilize TD-DFTB to investigate the optical properties of L-gln and L-pyro-amm within this approach.

\subsubsection{L-Pyroglutamine-Ammonium}

\begin{figure}[H]
  \centering
  \includegraphics[width=0.7\linewidth]{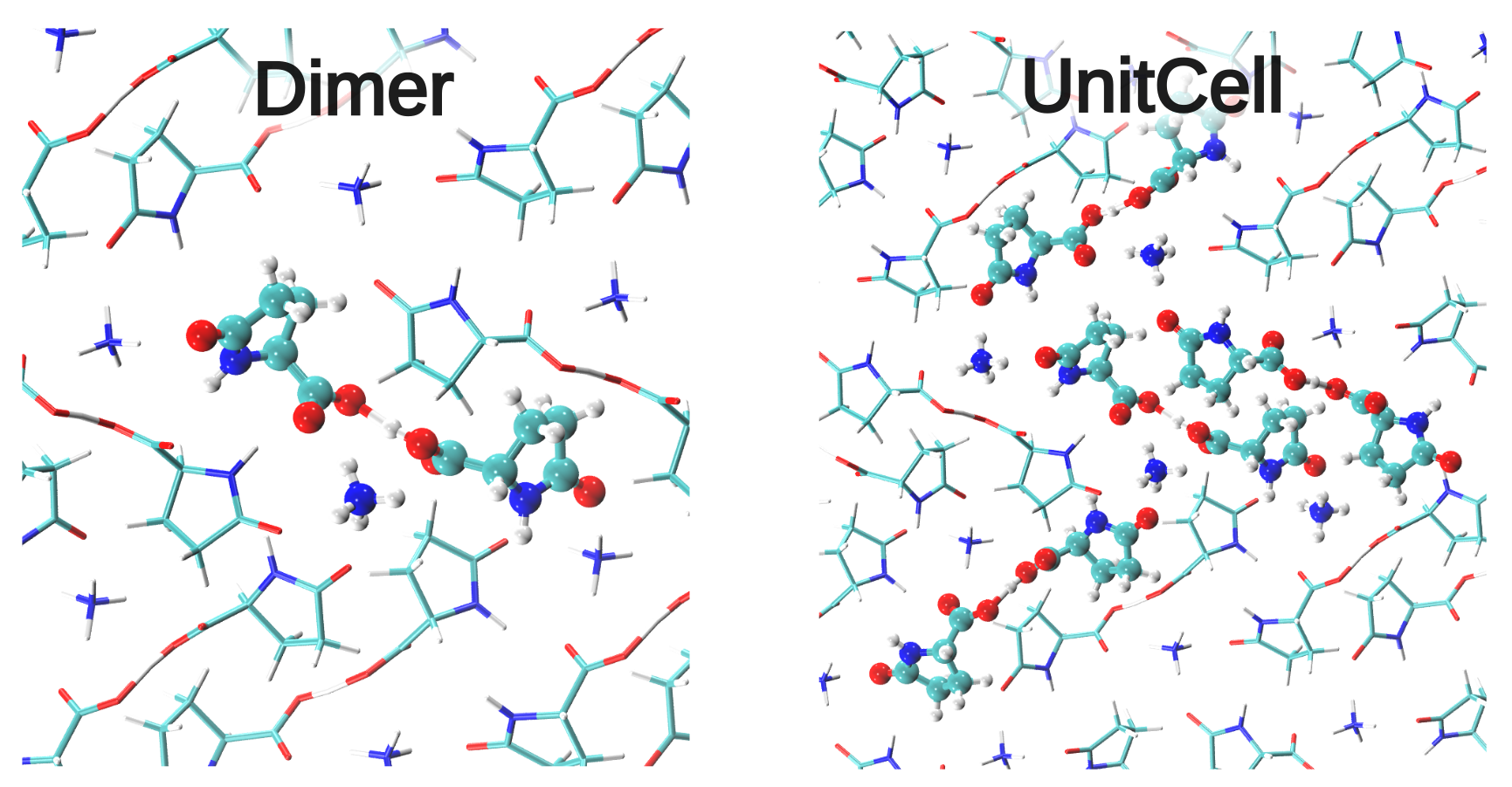}
  \caption{Dimer and Unit cell systems in the QM/MM simulations. The QM region is depicted in ball and stick while the MM in line representations. The real systems were replicated in the three directions; for the sake of clarity, here we are showing a 2D representation.}
  \label{fig:qmSys}
\end{figure}

We constructed two model QM regions embedded within an MM crystallographic environment including a dimer and the unit cell as shown in the Figure \ref{fig:qmSys}.

\begin{figure}[H]
  \centering
  \includegraphics[width=1.0\linewidth]{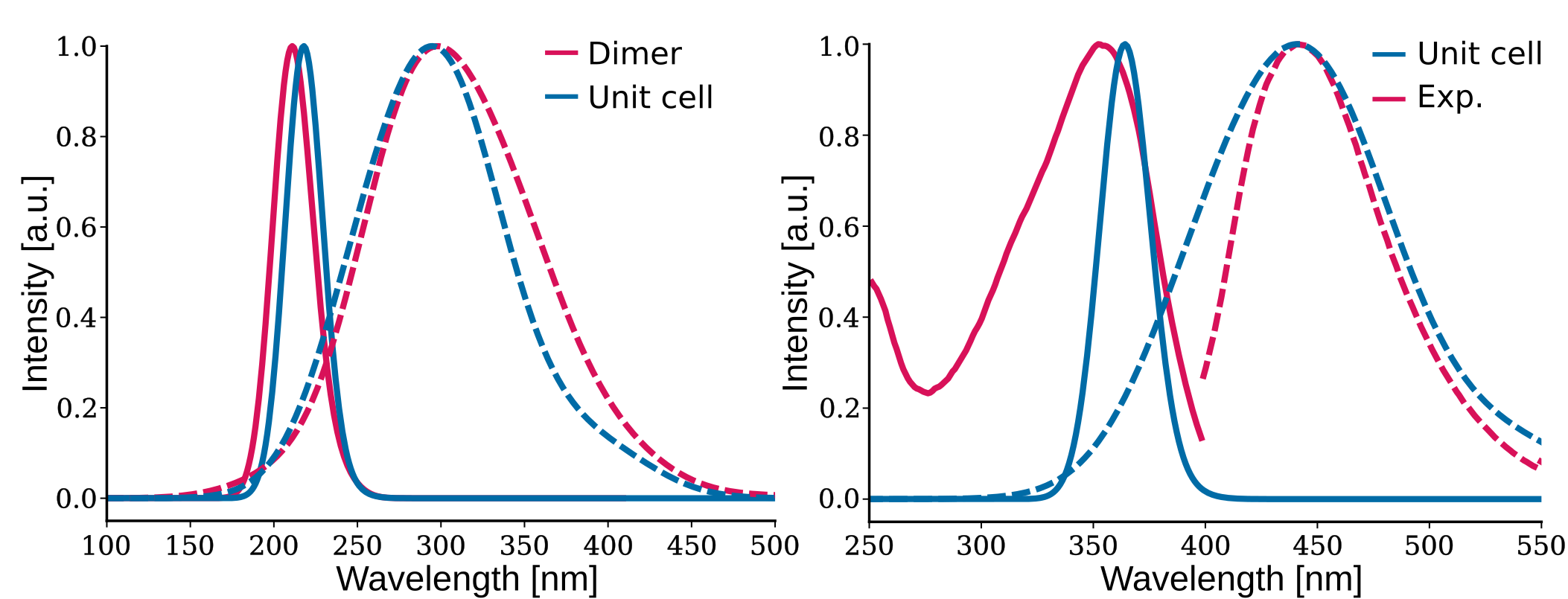}
  \caption{Excitation (solid lines) and Fluorescence (dashed lines) spectra for L-pyro-amm. Left panel shows the average spectra obtained from QM/MM simulations using the dimer (red line) and the unit cell (blue line) in the QM region. Right panel shows the experimental spectra (extracted from reference \cite{stephens2021short}, red line) and the average calculated from the QM/MM simulations for the unit cell in the QM region, where the calculated spectra were shifted to match the maximum in the experimental fluorescence spectra. All the spectra were normalized independently.}
  \label{fig:excfluComp}
\end{figure}

The experimental data used to compare the optical properties in this system was the excitation and emission spectra, as reported in ref \cite{stephens2021short}. It can be observed that the maximum of the experimental excitation spectrum is around 350 nm (red solid lines in the right panel in Figure \ref{fig:excfluComp}), which corresponds to the low lying transitions in the absorption spectrum in L-pyro-amm crystal (red solid line in the left panel of Figure \ref{fig:absPBC}). Therefore, to allow for a one-to-one comparison, we determined the excitation spectra instead of the absorption, \textit{i.e.}, including only the lower energy transitions in our calculations. We first decided to compute the excitation spectra using only the first transition ($S_0\to S_1$). Calculated and experimental results are shown in Figure \ref{fig:excfluComp}.

The left panel in Figure \ref{fig:excfluComp} shows the results for both QM systems in L-pyro-amm in the QM/MM simulation. We can observe that both systems show similar excitation maximum (approx. 215 nm) and emission (approx. 290 nm). When we compared these values with the experimental data (red lines in the right panel) we observe notable differences. This finding is a well known fact that TD-DFT and therefore in its approximate variant TD-DFTB, tend to exhibit a notable shift in the absolute magnitude of the transition energies when comparing with experimental measurements\cite{caricato2010electronic,leang2012benchmarking}. 
An additional factor contributing to the discrepancy between our computational outcomes and experimental findings lies in the inherent limitations of the QM/MM approach in representing the environmental effects. As demonstrated in Section \ref{sec:resultsPBC_abs}, our methodology achieves outstanding agreement with experimental results when employing a sampling coming from \textit{ab-initio} ground state simulations with periodic boundary conditions (see Figure \ref{fig:absPBC}).
Regrettably, this level of concordance is not replicated in our QM/MM simulations. To illustrate the impact of the MM environment, we conducted excited state calculations for the unit cell system using the same conformations employed for calculating the excitation spectra (blue solid line in the left panel of Figure \ref{fig:excfluComp}), with all MM atoms removed. The results are presented in Figure S4 in the Supplementary Information. We successfully reproduced the calculated absorption spectra depicted in Figure \ref{fig:absPBC} (red solid line in the right panel), thereby demonstrating the significant blue-shift effect induced by the MM environment. However, as we will elucidate in the following, these disparities persist not only in the ground state but also in excited state simulations. Consequently, focusing on energy differences becomes a more pertinent and meaningful avenue for comparison. We have thus determined the Stokes shift, defined as the difference between the maximum of the emission band and the absorption or excitation spectra the results of which are summarized in Table \ref{tbl:stokes}. Notably, we find remarkable agreement between the experimental results and the two model systems for the Stokes shift. It is also interesting to observe that the dimer system appears to fully capture the energetics associated with the fluorescence Stokes shift which reinforces the use of the dimer model systems in disentangling the non-radiative decay mechanisms.

\begin{table}[H]
\begin{center}
  \begin{tabular}{ c c c c }
    \hline
    System & Absorption/Emission Max. [nm] & FWHM$^{a}$ [nm] & Stokes shift [nm] \\
    \hline
    Dimer & 211/298 & 120 & 87 \\ 
    Unit Cell & 218/295 & 121 & 77 \\
    Experiment$^{b}$ & 354/441 & 92 & 87 \\
    \hline
  \end{tabular}
  \caption{Stokes shift for the different models in the QM/MM simulations}
  \label{tbl:stokes}
  \begin{tablenotes}
      \small
      \item $a $ FWHM denotes the Full Width at Half Maximum for the fluorescence spectra.
      \item $b $ Experimental results were extracted from the reference \cite{stephens2021short}.
  \end{tablenotes}
\end{center}
\end{table}

The right panel of Figure \ref{fig:excfluComp} presents the excitation and fluorescence spectra for both experimental and calculated using the unit cell of L-pyro-amm in the QM/MM simulations. The calculated spectra were adjusted to align with the peak of the emission spectra. Impressively, our calculations are in rather modest agreement with the experimental data. The deviation of 10 nm in the Stokes shift and also the slight shoulders at 320 nm present in the experimental excitation spectra (red solid line) which are absent in the emission (red dashed line) are an indication that the excitation spectra involves more than only one transition from the ground state.  Consequently, we computed the excitation spectra by incorporating an increasing number of transitions to accommodate this behavior. We have noted that incorporating additional states leads to a slight improvement in the agreement of the Stokes shift with the experimental data, up to the inclusion of 10 states. However, beyond this point, the results show poor agreement, and the full width at half maximum (FWHM) is also not accurately reproduced in all cases (see Figure S5 and Table S1 in the Supplementary Information).

\subsubsection{L-Glutamine}
\begin{figure}[H]
  \centering
  \includegraphics[width=0.8\linewidth]{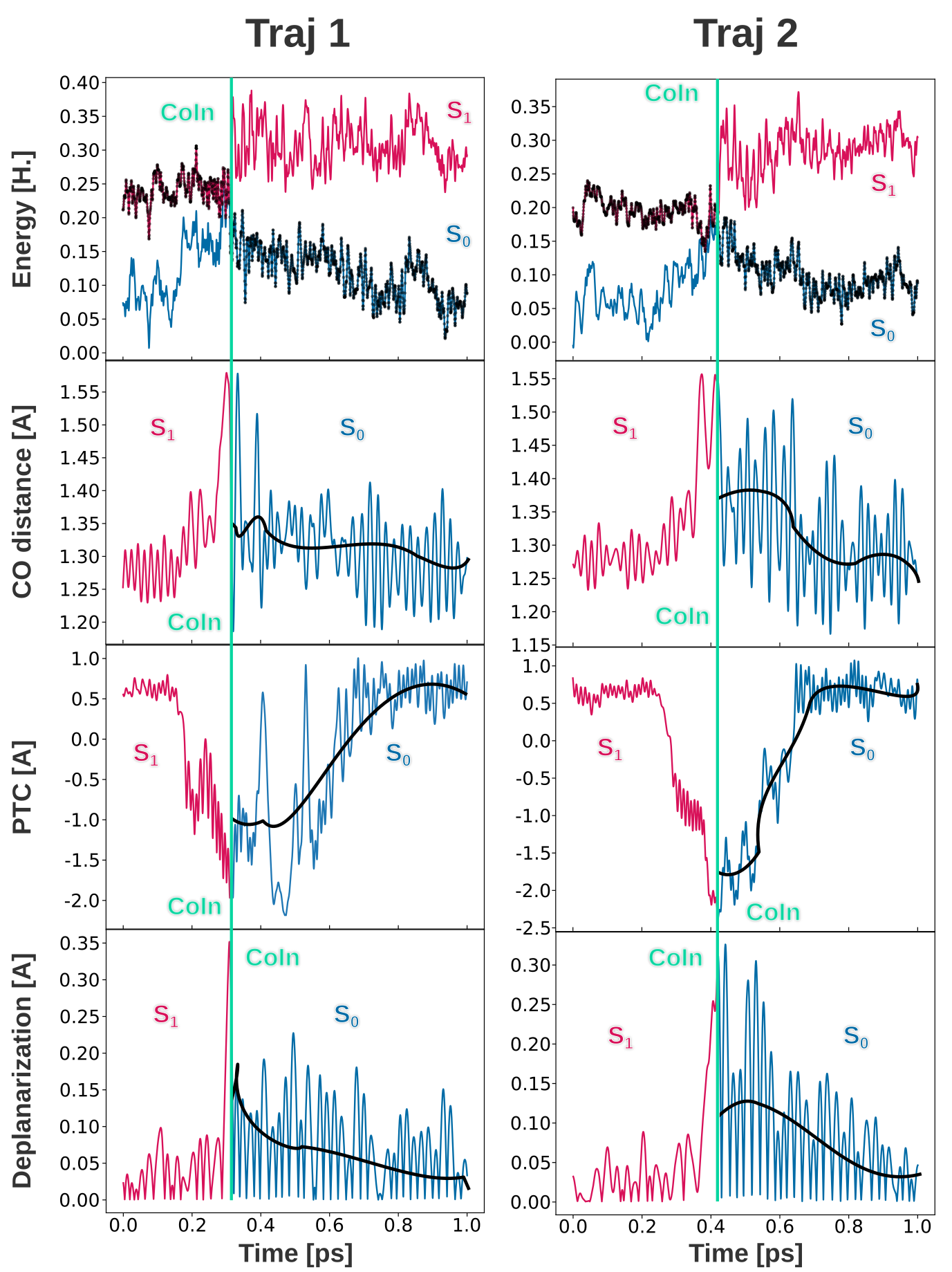}
  \caption{Temporal evolution of the most relevant features for two TSH trajectories in which we observed non-radiative decay in QM/MM simulation for the unit cell of L-gln system. Top panels show the evolution of the potential energies surfaces $S_1$ (red line) and $S_0$ (blue line), where the black dot indicates the current surface along the time. The other panels show the evolution of CO mode, Proton transfer Coordinate (PTC) and the Deplanarization mode as a function of time. Vertical green lines indicate the time of the transition $S_1\to S_0$ (CoIn). Black lines show the average of each modes after the CoIn.}
  \label{fig:modes_qmmm}
\end{figure}

In the case of L-gln, the non-fluorescent system, our objective was to investigate the non-radiative relaxation mechanism from the $S_1$ electronic state. It's important to note that the purpose of this section is to gain insights into the distinctions compared to the dimer in vacuum, rather than providing a quantitative analysis. Consequently, we conducted simulations for 10 trajectories. Figure \ref{fig:modes_qmmm} depicts the evolution of the potential energy surfaces ($S_1$ and $S_0$) along with the pertinent modes coupled to the CoIn, as discussed earlier, for two TSH trajectories in which we observed the non-radiative transition employing QM/MM simulations where the QM region is the unit cell.

From both trajectories, it is evident that the qualitative findings align with those discovered in the dimer case of L-gln in vacuum. Notable features include the stretching of the CO bond, the proton transfer and the deplanarization of the amide group in the proximity of the Conical Intersections (CoIn). It's worth noting that the characteristic lifetimes for this non-radiative decay are now longer in the QM/MM simulations, at 306 and 421 fs, as compared to the vacuum dimer (approximately 50 fs). This difference is likely due to the increased complexity in the unit cell system, offering more degrees of freedom and thus a broader range of relaxation pathways. However, as discussed previously, the photophysical mechanism governing the $S_1 \to S_0$ transition remains consistent across both the unit cell QM/MM and vacuum dimer scenarios. 

An interesting distinction arises after the system encounters the CoIn. As illustrated in Figure \ref{fig:modes_qmmm}, all relevant modes persist in an excited vibrational state on the $S_0$ surface. In our simulations, we chose to employ the NVE ensemble, avoiding the use of a thermostat. This decision was motivated by the fact that introducing a thermostat could potentially alter the velocities of the atoms, resulting in a different pattern of vibrational relaxation in the excited states. Such alterations might be attributed to the system's interaction with the thermal bath rather than its inherent photophysical behavior. Therefore, achieving a comprehensive understanding of vibrational relaxation would necessitate longer simulation times than the scope of this methodology allows.

\begin{figure}[H]
  \centering
  \includegraphics[width=0.8\linewidth]{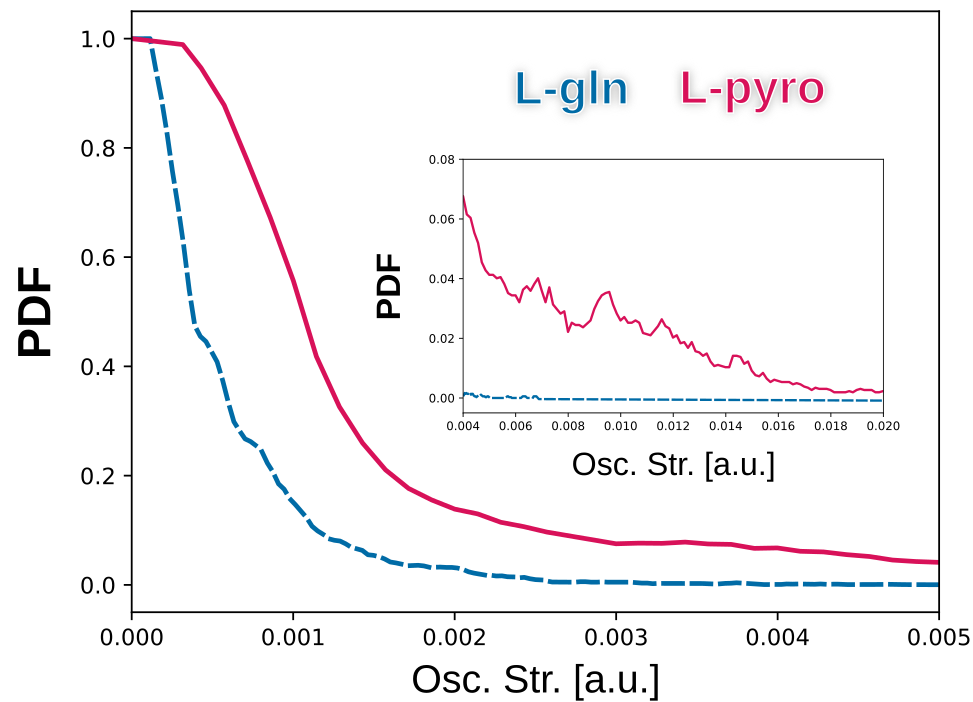}
  \caption{Oscillator strength distribution for the transition $S_1\to S_0$ for L-gln (blue dashed line) and L-pyro-amm (red solid line) systems, employing the unit cell in the QM/MM simulations. The inset shows the same distribution at higher oscillator strength.}
  \label{fig:fosc_qmmm}
\end{figure}

Up to this point, our primary emphasis regarding the fluorescence property has been centered on exploring the non-radiative decay within L-gln and L-pyro-amm, utilizing a wide variety of model systems and methodologies. The rationale behind this approach was to gain insights into the emission process, given the competitive nature of both radiative and non-radiative pathways. However, as previously discussed, not all the trajectories in L-gln resulted in a non-radiative decay to the ground state. Thus, an interesting point of comparison lies in the emission probabilities between the L-gln and L-pyro-amm systems. This involves calculating the distribution of oscillator strengths derived from TSH simulations, where non-radiative decay was not observed. Consequently, the system maintains its evolution on the $S_1$ potential energy surface. The resulting distribution is illustrated in Figure \ref{fig:fosc_qmmm}.

As we can discern from Figure \ref{fig:fosc_qmmm}, both compounds display a peak close to zero oscillator strength. This aligns with the fact that the L-gln system lacks fluorescence emission. In the case of L-pyro-amm, its fluorescence is relatively weak in intensity, several orders below the typical aromatic compounds\cite{morzan2022non}, yet it remains detectable through conventional spectroscopic methods. Furthermore, the plot reveals that as the oscillator strength increases, the probability of emission is notably higher for L-pyro-amm compared to L-gln, which strongly concurs with experimental findings\cite{stephens2021short}. Moreover, a minor population of L-pyro-amm conformations demonstrates significantly elevated oscillator strength values (refer to the inset in Figure \ref{fig:fosc_qmmm}). Although this population is minor, it is evidently greater than what's observed for the non-fluorescent L-gln system.

\section{Conclusions}
In this study, we explored the phenomenon of \textit{Non-Aromatic Fluorescence} (NAF) in L-pyro-amm and its comparison with the non fluorescent L-gln system, leveraging the computational efficiency offered by the density functional tight-binding (DFTB) method. Our application of DFTB yielded a robust structural and chemical description of the hydrogen bonds (HB) present in these structures. 
The correct behavior of the HBs in the L-gln system, as well as the characteristic double-well structure in short hydrogen bonds present in L-pyro-amm, was successfully reproduced. These properties were recognized as being pivotal for NAF across various systems.

The integration of trajectory surface hopping (TSH) with time-dependent DFTB in our study provided a mutual consensus of results with higher-level theories, exemplifying the versatility and efficacy of this combined approach. The computational efficiency inherent in TD-DFTB enabled us to attain additional statistical significance, particularly advantageous in comprehending the less-frequent non-radiative decay in the L-pyro-amm fluorescence system. Despite these strengths, subtle differences surfaced. Agreement in modes like carbonyl and proton transfer with our previous work highlighted the consistency of key features. On the other hand, modes such as the de-planarization display more differences compared to the previous time-dependent density functional theory findings. The precise origin of this observed difference could be an inherent limitation of the DFTB method or arise as a consequence of our non adiabatic dynamics approach, specifically the utilization of the Landau-Zener method (LZSH). In this method, the focus in determining non radiative probabilities is on the energy gap between the states. As soon as the system reaches a minimum energy gap, the algorithm initiates a transition between potential energy surfaces. In contrast, Fewest-Switches method (FSSH) incorporates non adiabatic coupling vectors, which significantly influence the hopping criteria, taking into account both the energy gap and the vectors. Consequently, our results exhibit a slightly accelerated decay to the ground state with LZSH compared to those in FSSH, reducing the likelihood of observing the minor contribution from the deplanarization mode. Clearly, this discrepancy necessitates further investigation in future studies.

In addition, our investigation extended to larger systems within the QM regions and incorporating environmental effects through the QM/MM approach, reveals that the absolute values for the QM/MM absorption spectra do not precisely match experimental observations. This discrepancy can potentially be mitigated by employing MM polarizable models\cite{li2015polarizable} or increasing the size of the QM region\cite{kulik2016large,isborn2012electronic}. Despite these differences, our results present a consistent non-radiative mechanism in the L-gln system. Futhermore, the scalability of our simulations to encompass more degrees of freedom in a larger system elucidates that the observed decay mechanisms persist but manifest over extended time. The computational efficiency afforded by DFTB method proves indispensable in capturing these prolonged processes, which would be impractical with traditional ab-initio methods. 

\section{Associated Content}
Supporting Information Available:\hfill \break Systems models employed in non-adiabatic dynamics in vacuum; full range of the absorption spectra for both crystal systems; molecular orbitals involved in the first excitation; environmental effects on the absorption spectrum; excitation spectra with increasing number of excitations.\hfill \break This information is available free of charge at the website: \url{http://pubs.acs.org/}

\begin{acknowledgement}

The authors would like to express their gratitude to Bálint Aradi for engaging in insightful discussions and giving feedback regarding the research. GDM, DB and AH thank to the European Commission for funding on the ERC Grant HyBOP 101043272. GDM, UNM and AH also acknowledges to CINECA supercomputing for the resource allocation (project NAFAA-HP10B4ZBB2 and V-CoIns-HP10BY0AET). CRL-M and MAS acknowledge financial support from the German Research Foundation (DFG) through Grant No. FR 2833/82-1. CRL-M extends special thanks to UNM for the invitation to the "Workshop on Frontiers in Excited State Electronic Structure Methods: from Spectroscopy to Photochemistry" last year in Trieste, which served as the catalyst for initiating this collaboration between our groups.

\end{acknowledgement}
\providecommand{\latin}[1]{#1}
\makeatletter
\providecommand{\doi}
  {\begingroup\let\do\@makeother\dospecials
  \catcode`\{=1 \catcode`\}=2 \doi@aux}
\providecommand{\doi@aux}[1]{\endgroup\texttt{#1}}
\makeatother
\providecommand*\mcitethebibliography{\thebibliography}
\csname @ifundefined\endcsname{endmcitethebibliography}
  {\let\endmcitethebibliography\endthebibliography}{}

\end{document}